\documentstyle[epsf,prd,aps,floats]{revtex}
\newcommand{\blankline}{\vskip .3cm}
\newcommand{\f}{\begin{equation}}
\newcommand{\ff}{\end{equation}}
\newcommand{\be}{\begin{equation}}
\newcommand{\ee}{\end{equation}}
\newcommand{\bea}{\begin{eqnarray}}
\newcommand{\eea}{\end{eqnarray}}

\begin{document}
\twocolumn[\hsize\textwidth\columnwidth\hsize\csname@twocolumnfalse\endcsname  

\title{ Non-Linear Relativity in Position Space}
\author{Dagny Kimberly, Jo\~ao
Magueijo and Jo\~ao Medeiros} 
\address{\ The
Blackett Laboratory, Imperial College of Science, Technology and
Medicine } \blankline \blankline \blankline \blankline
\maketitle

\begin{abstract}
We propose two methods for obtaining the position space
of non-linear relativity, i.e. the dual of its usual 
momentum space formulation. In the first approach we require that 
plane waves still be solutions to
free field theory. This is equivalent to postulating the invariance
of the linear contraction between position and momentum spaces, and dictates a 
set of energy-dependent spacetime Lorentz transformations. In
turn this leads to an energy dependent metric. The
second, more problematic approach allows for the position 
space to acquire a non-linear representation of the Lorentz group 
independently of the chosen representation in momentum space. 
This requires a non-linear contraction 
between momentum and position spaces. 
We discuss a variety of physical implications of these approaches
and show how they point to two rather distinct formulations of 
theories of gravity with an invariant energy and/or
length scale. \end{abstract}
\pacs{PACS Numbers: *** }
 ]
\renewcommand{\thefootnote}{\arabic{footnote}}
\setcounter{footnote}{0}

\section{Introduction}
A number of paradoxes in quantum gravity (e.g. \cite{leejoao}) suggest that 
it may be desirable to re-examine the nature of Lorentz invariance.
We briefly review one example. We know that 
the combination of gravity ($G$), the quantum ($\hbar$) and 
relativity ($c$) gives rise to the Planck length, 
$l_P = \sqrt{\hbar G / c^3 } $, the Planck
time $t_P=l_P/c$, and the Planck energy $E_P= h/t_P$. These scales
mark thresholds beyond which the classical description of
space-time breaks down and qualitatively new phenomena are
expected to appear. However, whatever quantum gravity  may turn out 
to be, it is expected to agree with relativity
for all experiments probing the nature of space-time at, say, length 
scales much larger than $l_P$. So the question arises: in whose
reference frame are $l_P$, $t_P$ and $E_P$ the thresholds for new
phenomena? For suppose that there is a physical length scale
which measures the size of spatial structures in quantum
space-times, such as the discrete area and volume predicted by
loop quantum gravity. Then, if this scale is $l_P$ in one inertial
reference frame, special relativity suggests it may be different
in another observer's frame -- a straightforward implication of
Lorentz-Fitzgerald contraction.

Solutions to these problems have been proposed within
the framework of what is variously known as  non-linear, deformed, 
or doubly special relativity (DSR)~\cite{amelstat,gli,leejoao}. 
These theories may be understood as non-linear realizations 
of the Lorentz group~\cite{leejoao1}, in which modified Lorentz 
transformations reduce to the usual ones at low energies, but
leave invariant $E_P$, the borderline between classical and quantum gravity. 
These theories can explain the recent claims for 
high-energy cosmic ray anomalies~\cite{review,crexp,cosmicray,leejoao1}
(see also~\cite{amel,amel1,liouv}). This is due to 
the introduction of deformed dispersion relations 
(i.e. departures from $E^2-p^2=m^2$). Even though considerable
doubts still hang over these results (and considerable caution should
therefore be exercised by theorists), they are intriguing, and
if true, very far-reaching indeed.

Given that deformed dispersion relations provide for observational input
into these theories, a momentum space formulation has been
preferred in the past (see~\cite{fock,man,step} for notable exceptions).
Once this choice is made, however, recovering position space 
is highly non-trivial. With loss of
linearity, duals no longer mimic one another, i.e. vectors no
longer transform according to the inverse linear transformation
of co-vectors. Defining the spacetime structure of DSR
is therefore highly non-trivial.

This is not merely a formal annoyance, as the physical
interpretation of many results obtained in momentum space can only
be fully assessed when the relation to position space is
clarified. For instance, as recently discussed \cite{vel1,vel2,vel3},
one may have trouble defining physical velocity in these theories.
The primary definition of velocity should be the ``space-time''
velocity: \be v={dx\over dt}, \ee which is the relevant
quantity, for example, in gamma-ray timing experiments \cite{gama}. 
It has been
argued that the Hamiltonian expression for the velocity,
\be\label{ham} v={dE\over dp} \ee should be preserved, as this is
the group velocity in normal circumstances. However, this has
never been generally proved (with \cite{vel1} even claiming that
this leads to contradictions; see also \cite{vel3}). 
Clearly this matter can only be
settled when one rediscovers position space from the usual
momentum space formulation, and the answer {\it will depend} on
the particular construction of position space employed. (This
matter only adds complexity to the issues raised
in~\cite{toller1,Grum,ahl}).

In this paper we propose two possible ways to rediscover
position space in non-linear relativity. 
The first approach might seem the least obvious, but turns out to 
be more rewarding in terms of physical implications
and internal consistency. It should be seen as the heart of this paper. 
The second approach mirrors techniques used for setting up DSR in 
momentum space, but runs into severe problems. We spell them out in 
the hope that they might be soluble.

In the first approach we start by requiring that free field theories 
have plane wave solutions, with momenta satisfying the set of deformed 
dispersion relations characterizing a given DSR theory. Even though
in Section~\ref{field}
we give explicit examples of such field theories, our paper does
not depend on their details, but only on the existence of plane
wave solutions. That such solutions exist is equivalent to requiring 
that the contraction between momenta and positions remain linear 
(a linear contraction provides the invariant phase for plane waves).

In Section~\ref{dual}
it is found that the invariance of such a linear contraction
fully fixes the form of the Lorentz transformations in position space.
Our main result is that the new Lorentz transformations are energy 
dependent, and so the space-time metric becomes energy dependent too.
The physical meaning of $E$ in this dependence is fully spelled out
in Section~\ref{example}. Further physical implications are explored
in Section~\ref{physics}.

The second approach, developed in Section~\ref{dual2},
is more unimaginative: we simply introduce
a non-linear representation of the Lorentz group in position
space. This mimics techniques usually used in momentum
space, and may be used to remove from the physical space 
durations smaller than $t_P$, solving the paradox which opened
this paper. However this approach also has
many undesirable features, for instance the contraction between 
position and momenta coordinates must assume a non-linear form. 
Even though we believe this approach is likely to run into
great difficulties we show how, historically, the first non-linear
realization of the Lorentz group was proposed within this
framework~\cite{fock,man,step}.

In this paper we shall set ``c'' (the low-energy or long-distance 
limit of the speed of light) equal to one, and use
metric signature $(-,+,+,+)$.

\section{Field theory realization of doubly special relativity}
\label{field}
In the usual formulation of DSR one starts from a set of deformed
dispersion  relations:
\be E^2\label{disp}
f^2 (E) - {\mathbf p}^2 g^2(E) = m^2, \ee
to set up a non-linear momentum space.
Except in trivial cases, relations (\ref{disp}) are not invariant under the 
standard, linear Lorentz transformations. Rather, they only comply with the
principle of relativity if we define a non-linear representation 
of the  Lorentz group via 
\be
K^i = U^{-1}  L_0^{\ i} U 
\label{newL}
\ee
where the $U$ map is given by 
\be \label{U}U_a (E , p_i) = (U_0,U_i)=\left(Ef(E)
,p_i g(E)\right) \ee  
and $ L_0^{\ i}$ are the usual Lorentz generators acting
on momentum. The dispersion relations (\ref{disp}) and the 
map $U$ given by (\ref{U}) are equivalent, and the 
$U$ map fully specifies the single particle sector of a given 
DSR formulated in momentum space. Several proposals can be found 
in the literature. 

We wish to propose an associated position space 
in such a way that the momentum space formulation uniquely fixes 
the transformation laws in position space. This can be achieved
by introducing an appropriately strong physical requirement;
for example by demanding that  there be plane wave solutions
to free field theories.  By this
we mean solutions of the form \be \phi=Ae^{-ip_a  x^a} \ee
where $p_a$ satisfies deformed dispersion
relations (\ref{disp}), and the contraction $p_a  x^a$ 
remains  linear ($p_a  x^a=p_0x^0 + p_ix^i$). Without a linear 
contraction such waves would not be ``plane''.

Even though this is the only postulate used in this approach,
it is possible to exhibit field theories satisfying this requirement.
Considering, say, scalar field theory, one should set up equations
of motion via 
the replacement
\be \label{presc} p_a\rightarrow i\partial_a \ee
applied to the dispersion relations (\ref{disp}). 
For example, if the momentum space
formulation is based upon the invariant:
\be\label{disp1}
{E^2-{\mathbf p}^2\over 1- (l_P p_0)^2}=m^2
\ee
we should have a ``Klein-Gordon'' equation:
\be
\left(\eta^{ab}{\partial_a\over 1-il_p\partial_0}{\partial_b\over
1+il_p\partial_0}+m^2\right)\phi=0 ,\ee 
This equation  is linear in $\phi$,
and has plane wave solutions $\phi=Ae^{-i p_a x^a}$
with $p_a$ satisfying the dispersion relations (\ref{disp1}).
It can be obtained from the Lagrangian:
\be{\cal L}={{1\over 2}\left[
\eta^{ab}{\partial_a\over 1+il_p\partial_0}\phi{\partial_b\over
1+il_p\partial_0}\phi +m^2 \phi^{2}\right]}.\ee
Another example is the $U$ chosen in \cite{leejoao}, for which
the deformed dispersion relations are:
\be\label{inv0} {\eta^{ab}p_ap_b\over
(1-l{_p}p_0)^2}=m^2 .\ee
In this case prescription (\ref{presc}) 
leads to 
\be
\left(\eta^{ab}{\partial_a\over 1-il_p\partial_0}{\partial_b\over
1-il_p\partial_0}+m^2\right)\phi=0 \, .\ee
Again there are plane wave solutions  with $p_a$ satisfying
(\ref{inv0}).

These examples are given as illustration, and the rest 
of our paper does not depend on the details of DSR field theory.
However we believe that this framework for relating deformed
dispersion relations and field theory is likely to be the field theory 
extension of DSR. DSR  was initially set up on 
purely kinematical grounds, considering classical point particles. 
With prescription (\ref{presc}) the field equations are always linear and 
have  plane wave solutions or their most general superposition. 
The plane waves carry a momentum satisying deformed dispersion
relations. Even though we only considered scalar fields it 
is possible to generalize our construction to arbitrary spins. 
The result should  be compared with the frameworks of~\cite{ahl,arzano}
and we shall comment on this matter at the end of the next section.

If the deformed dispersion relations 
have a maximum energy (as is the case with (\ref{inv0})), such field theories 
have a natural ``Lorentz invariant'' energy cut-off, and so 
are  finite when interactions are considered. Hence the 
issue of their renormalizability should not arise, since they
are naturally finite. A more concrete
example of the effect of deformed dispersion relations 
in effective field theory may be found in~\cite{myers}.

Finally, despite their aspect, the theories we have displayed
are ``local'' and ``causal''; clearly not in the sense of the linear 
Lorentz group, but in the modified sense of non-linear Lorentz 
transformations, which we shall now spell out.

\section{The dual space-time and its metric}\label{dual}

It turns out that the criterium used in the previous section for setting
up field theory is sufficient to fully fix the properties of 
position space. 
If there are plane wave solutions for scalar fields, then the linear
contraction $p_a x^a$ providing their phase must also be a
scalar. This fixes the form of the space-time Lorentz transformations, 
but now these are energy dependent.

Let us consider the general dispersion relations (\ref{disp})
and associated $U$ map (\ref{U}). Expression (\ref{newL}) shows
that $U$ is a linearizing map, taking physical momenta
into momenta which can be processed by the linear Lorentz generators
$L_0^i$. We would like to find a similar map for the $x$ coordinates,
i.e. a map that transforms $x^a$ into coordinates $U^a(x)$ which can
be transformed using the linear Lorentz transformations.  
$U^a(x)$ must contract linearly with $U_a(p)$. The fact that we require
the physical $x$ and $p$ to also contract linearly implies that:
\be U_a(p)U^a(x)=p_a x^a \ee
Therefore the linearizing map acting on position coordinates must
be:
\be \label{Udx} U^a (x) = (U^0,U^i)= \left({t\over f(E)} ,{x^i\over g(E)}
\right) \, . \ee
The deformed Lorentz transformations
acting upon $x$ can now be obtained from linear transformations
acting upon $U(x)$. They are: \bea \label {transf2}
t'&=&\gamma{\left(t-vx {f(E)\over g(E)}\right)}
{f(E')\over f(E)}\nonumber \\
x'&=&\gamma{\left(x-vt {g(E)\over f(E)}\right)} {g(E')\over
g(E) }\nonumber \\
y'&=&y{g(E')\over
g(E) }\nonumber \\
z'&=&z{g(E')\over g(E) } ,\eea where $E'$ is the Lorentz boosted
energy. Hence the space-time Lorentz transformations now depend
upon the energy, and space-time has become intrinsically mixed
with energy-momentum.

Transformations (\ref{transf2}) have the invariant:
\be \label{metric}{s^2}={\eta_{ab} U^a(x)U^b(x)}
= - {t^2\over f^2} + {(x^i)^2\over g^2}, \ee
Hence the empty space metric is now
\be
g_{ab}(E)={\rm diag} \left[ -{1\over f^{2}(E)}, {1\over g^{2}(E)}, 
{1\over g^{2}(E)}, {1\over g^{2}(E)}
\right]
\ee
In order to preserve $g_{ab} g^{bc}= \delta_a ^c$ we
should also have the inverse metric:
\be \label{metup}
g^{ab}(E)={\rm diag} [- f^2(E), g^2(E), g^2(E),g^2(E)]
\ee
Notice that in the low energy limit ($El_P \ll 1$)
we have $g_{\mu\nu}\approx \eta_{\mu\nu}$
and $g^{\mu\nu}\approx \eta^{\mu\nu}$.

This structure should elucidate the difference between the field theories 
proposed in the previous section, and those of~\cite{ahl,arzano}.
It was suggested in~\cite{ahl} that the non-linearities of DSR
could -- and should -- be undone when setting up field theory.
This was shown for fermionic fields first, but also claimed
for all spins (including, presumably, those of spin 0, i.e. scalar
fields). The argument
centered on the absence of a well defined position space dual to the 
non-linear momentum variables.

This issue was first solved by~\cite{arzano} with the introduction of 
non-commutative position space. In this paper we have introduced 
instead an energy-dependent position space. Field theories based
on these two approaches are quite distinct, and also crucially
different from those of~\cite{ahl}. But both stress
the non-trivial nature of DSR once a suitably modified position
space is introduced~\footnote{The argument that 
the non-linearities in DSR can be removed by making a redefinition 
of energy and momenta has been refuted by two further developments:
the explict example of 2+1 
gravity~\cite{2p1} and the realization that DSR's phase space
has an {\it invariant} curvature~\cite{desit,moffatcurvp}.}.

\section{An example and some physical considerations}\label{example}

As an example we may consider the form of $U$ chosen in \cite{leejoao},
associated with dispersion relations (\ref{inv0}).  In this case it 
was found that energy and momentum boosts in the $x$-direction 
assume the form
\bea
E'&=&{\gamma(E-vp_x)\over (1+(\gamma-1)l_pE-\gamma l_pvp_x)}\nonumber\\
p'{_x}&=&{\gamma(p_x-vE)\over (1+(\gamma-1)l_pE-\gamma l_pvp_x )}\nonumber\\
p'{_y}&=&{p_y\over (1+(\gamma-1)l_pE-\gamma l_pvp_x )}\nonumber\\
p'{_z}&=&{p_z\over (1+(\gamma-1)l_pE-\gamma l_pvp_x )}
\eea
The energy-momentum invariant is (\ref{inv0}), from which
one can read off the forms of the functions $f$ and $g$. Thus we find
that \bea
t'&=&\gamma(t-vx)[1+(\gamma-1)l_pE-\gamma l_pvp_x]\nonumber\\
x'&=&\gamma(x-vt)[1+(\gamma-1)l_pE-\gamma l_pvp_x]\nonumber\\
y'&=&y[1+(\gamma-1)l_pE-\gamma l_pvp_x]\nonumber\\
z'&=&z[1+(\gamma-1)l_pE-\gamma l_pvp_x]\label{lorex1}
\eea
Transformations (\ref{transf2}) have the invariant:
\be {s^2}={-t^2+(x^i)^2\over (1-l_pE)^2}
\ee

Faced with energy dependent space-time transformations
and metric a natural issue is the physical meaning of $E$ 
in these transformations. Whose $E$ is this?  
To answer this question we note that 
our construction follows from demanding that 
free field theories should accept plane wave solutions. In turn
this requires that the linear contraction providing their phases should be a
scalar or invariant. Thus the $E$ appearing in the Lorentz 
transformations is the $E$ of the plane wave used
to probe space-time. 

A similar argument can be made for wave packets and 
``particles''. The transformation laws for the position of a given 
particle depend on its energy as seen by a given observer.
Particles with  different energy, at the same position and time,
must transform differently and feel a different metric.
Different observers see a given particle with different energies,
and so assign to it different Lorentz transformations and metric.

So not only different observers see a given particle being affected
by different metrics, but the same observer will assign different 
metrics to different particles. The whole construction is clearly
not invariant in the sense of linear Lorentz transformations --
it is invariant (by construction) in the sense of non-linear momentum
Lorentz transformations .

In our picture space-time and energy and momentum have become intertwined.
Even ignoring gravitation, we cannot talk about time and position 
without a particle being there. The properties of this particle's position 
and time depend on its energy. At low energies, the dependence is very weak, 
so we have the illusion that space-time exists independently of the 
(test) particles that might fill them.

\section{Further physical implications}\label{physics}

Lorentz contraction and time dilation are quite different
in this theory. Let us consider first the example of 
transformations (\ref{lorex1}). 
If we boost a stationary rod with one endpoint attached to the origin 
and the other at
$x=l$, we find that its tip transforms into $x'=l'-vt'$, with
\be
l'={l\over \gamma}[1+(\gamma-1)l_pE_0].
\ee
where
\be
E_0={m\over 1+l_Pm} 
\ee
is the particle's rest energy (see~\cite{leejoao1}; also
Eq.~(\ref{inv0})). This provides the revised Lorentz 
contraction formula, and we notice that for $E_0=E_p$ all lengths
are invariant: $l=l'$. From $x'=l'-vt'$ we see that the particle's velocity
in the new frame is $v$. This  shows that the boost parameter $v$ is
the origin's velocity, something far from obvious as discussed
in the literature. 

Boosting $x=v_0t$ we may also show that the
law of addition of velocities is the same as in special relativity:
\be
v'={v-v_0\over 1-vv_0}
\ee
The time dilation formula, on the other hand, becomes:
\be
\Delta t'=\gamma \Delta t [1+(\gamma-1)l_pE_0]\, .
\ee
More generally for the extreme case of a Planck energy wave
($p_0l_P=1$), at rest we have $p_{a}=(E_p,0,0,0)$, so that
\bea
t'&=&\gamma^2(t-vx)\\
x'&=&\gamma^2(x-vt)\\
y'&=&y\gamma\\
z'&=&z\gamma
\eea 
As $v\rightarrow c$, $t'$ and $x'$ approach infinity faster
than regular Lorentz boosts by a factor of $\gamma$. 
There is also a transverse
effect, absent in the usual theory.

Performing a similar exercise with the more general transformation
(\ref{transf2}) we find the time dilation formula
\be \Delta t'=\gamma \Delta t {f(E')\over
f(E)}\ee and the Lorentz contraction formula: \be 
l'={l \over \gamma}{g(E')\over g(E)}\ee 
In general the boost parameter $v$ is not the velocity,
of the moving frame, $v_f$, which is
\be
v_f=v{g(E')\over f(E')}
\ee
By transforming $x=vt$ we would also arrive at the general 
law of addition of velocities. However, the result
is at first cumbersome --  and shall not be reproduced here -- 
until we consider the following.

Setting $m=0$ in the deformed dispersion relations (\ref{disp}), 
we find that the coordinate speed of massless plane waves (with phase
$x^ap_a=Et-{\bf p}\cdot {\bf x}$) is
\be \label{sol}c(E)={dx\over dt}={E\over p}={g(E)\over f(E)}.\ee 
This is  in
contradiction with the usual Hamiltonian expression \be c={dE\over
dp}. \ee Thus, in this formalism we have an energy-dependent speed
of light, but its speed is given by $E/p$, not $dE/dp$. This is
also the speed of null signals, obtained by setting $ds^2=0$ 
in Eq.~(\ref{metric}). Thus massless waves move along null 
surfaces. As we shall see now, the speed of light 
is also an observer invariant
speed in a suitably modified sense.

Given (\ref{sol}), the general case of the law of 
addition of velocities may be written in a more amenabel form:
\be
\frac{v'}{c(E')}={{v_0\over c(E)}-{v_f\over c(E')}
\over 1- {v_0\over c(E)}{v_f\over c(E')}}
\ee
This shows that the law $c=c(E)$ is an invariant. If light rays
in one frame propagate with $c(E)$, then in another frame their
energy will be $E'$ and their transformed speed will be $c(E')$.

We finally note that we have not considered here wave packets,
but it's conceivable that their ``space-time'' speed differs 
from the expression above. This has the implication that the 
space-time Lorentz transformations for wave packets may be 
different than those for plane waves.

\section{A less promising approach}\label{dual2}

It was shown in  \cite{leejoao1} how to introduce a general non-linear
representation of the action of the Lorentz group in momentum
space, corresponding to given deformed dispersion relations. A
formally identical procedure can be applied in position space. The
non-linearity that arises in doing so destroys translational
invariance.  Thus, all considerations in this section are to be
seen as statements made in reference to small displacements about
the origin (which can be a generic point), 
in the same sense that free falling frames in general
relativity correspond to local neighbourhoods.
The position space in this section is thus comprised of a set of 
coordinates $dx^a$ rather than global coordinates $x^a$.
Because they are all differences {\it with respect to the origin}
they should not be added or subtracted. Indeed the non-linearity
of their transformation would preclude such an operation (one
could, of course add and subtract these quantities non-linearly
and thus achieve covariance, but that is not necessary here).

In analogy with \cite{leejoao1} we construct non-linear realizations of
the Lorentz group in position space by means of a
transformation $V(x)$ such that the new generators of the Lorentz
group with respect to space-time coordinates are
\begin{equation}\label{V}
K^i = V^{-1} L_0^{\ i} V,
\end{equation}
where \be L_{ab} = x_a {\partial \over \partial x^b} - x_b
{\partial \over \partial x^a} \ee are the standard Lorentz
generators. Exponentiation of these generators reveals a
non-linear realization of the group.
Naturally this requires that
the contraction between infinitesimal displacements (position)
and momentum be non-linear, for it to remain
invariant. The contraction should be defined as
\be\label{contr}
<p, x>=U_a(p)V^a(x)\; ,
\ee
Sadly, this implies the absence even locally of plane wave solutions,
and thus of a suitable Hilbert space.

The first non-linear representation of the Lorentz 
group~\cite{fock,man,step} to have been proposed precedes
DSR and is an example of this procedure.
Taking\be\label{V2}
V^a(x)=x^a{1 \over t + R/c_0}, \ee leads to: \bea
t'&=&{\gamma\left(t- v x \right) \over
1-(\gamma -1){t\over R} +\gamma{v x
\over R}}\\
x'&=&{\gamma\left(x- v
t\right) \over 1-(\gamma -1){t\over R} +\gamma{v x
\over R}}\\
y'&=&{y \over 1-(\gamma -1){t\over R}
+\gamma{v x \over R}} \\
z'&=&{z \over 1-(\gamma -1){t\over R}
+\gamma{v x \over R}}  ,\eea 
the transformation first proposed by Fock. This transformation
produces an invariant length $R$, but given that 
$R$ is large instead of small this
transformation doesn't seem to play a role in quantum gravity. We
now discuss conditions upon $V$ leading to more interesting
transformations.

Whereas the pure momentum space formulation of non-linear relativity
treats the Planck energy $E_p$ as an invariant, we now require that the Planck
time $t_p$ be an invariant.  The only invariant times in
{\it linear} relativity occur at zero and infinity, which implies
for the {\it non-linear} case that $V(t_P)=0$ or $V(t_P)=\infty$
should hold. However, we want the range of the transformed time
coordinates, $V(t)$, to cover the domain of the boost
transformations for all values of the boost parameter $v$. 
Thus, the range of $V(t)$ should be
$[0,\infty]$.  In addition we need $V\approx 1$ in the $t\gg t_P$
limit such that the ordinary Lorentz transformations are recovered
for large $t$. Combining these conditions implies that $V(t_P)=0$
is the correct condition. (Note that in momentum space, on the
other hand, the appropriate condition is $U(E_P)=\infty$; this is
discussed in detail in~\cite{leejoao1}).

The simplest theory we found with these properties is
generated by the transformation
 \be\label{V1}
V^a(x)=x^a{\left(1-{t_P\over t}\right)}, \ee which
can be written as the exponential operator \be V=e^{-{t\over
t_P}D}, \ee where $D=x^a\partial_a$ is a dilatation. The boost
generators then take the form \be K^i=L_0^i-{t_Px^i\over t^2}D
,\ee which, in turn, induce a non-linear representation of a
Lorentz boost in the $x$-direction given by \bea
t'&=&\gamma (t-vx) {\left(1-{t_P\over t}\right)} +t_P\label{tboost}\\
x' &=&\gamma (x -vt){\left(1-{t_P\over t}
+ {t_P\over \gamma(t-vx)}\right)} \label{xboost}\\
y' &=&y{\left(1-{t_P\over t}
+ {t_P\over \gamma(t-vx)}\right)} \\
z' &=& z{\left(1-{t_P\over t}
+ {t_P\over \gamma(t-vx)}\right)} 
\eea 
From (\ref{tboost}) we see that 
the time dilation formula is now
\be \Delta t'=\gamma{\left(\Delta t- t_P \right)}
+t_P. \ee 
It follows that the duration $\Delta t= t_P$ is invariant,
as expected.  We may now invariantly impose as a physical condition 
$\Delta t>t_P$.  Hence we may remove sub-Planckian durations
from the theory in an invariant form, solving the paradox 
that opened this paper. Observers will never mix sub and super
Planckian regimes.

The transformation just proposed may be refined in a variety
of ways. In the proposed transformation we have broken time reversal
invariance. This may be restored by choosing
\bea\label{V3}
V^a(x)&=&x^a{\left(1-{t_P\over t}\right)} \quad{\rm  for }\quad
t>t_P\nonumber\\
&=& x^a{\left(1+{t_P\over t}\right)}\quad {\rm  for }\quad
t<-t_P\nonumber\\
&=&0\quad \quad\quad \quad\quad\quad{\rm      otherwise }
\eea
Then the Lorentz transformations derived above are valid for
$t>t_P$ (which implies $t'>t_P$). For $t<-t_P$ one should 
swap the sign of $t_P$. Durations $-t_P<t<t_P$ may now be 
removed invariantly. If we want to remove sub-Planckian
durations without having to posit an (invariant)
physical conditions we could
choose:
\be V=\log {\left(1-{\left|t_P\over t\right|}\right)}.\ee
The range $-t_P<t<t_P$ does not have an image inside the
resulting representation of the Lorentz group, and so is 
removed without the need to impose any supplementary condition.

If time durations smaller than $t_P$ become non-physical,
the concept of exact simultaneity becomes meaningless.
Concomitantly, lengths (which are spatial separations evaluated at
the same time) depend on how we redefine simultaneity.
Thus, instead of deriving the Lorentz contraction
formula with respect to exact simultaneity, we let $t$ represent a
fixed course-grained duration $T$, to which we refer length measures.
The resulting Lorentz contraction formula, obtained from
(\ref{tboost}) and (\ref{xboost}) with this prescription,
does not have a simple analytic form, but is straightforward to
implement numerically. Naturally $T\ll l$ but $T\ge t_P$, so that
lengths $l$ smaller than $l_P$ also become unphysical in this theory.

Finally note that the finite space-time invariant, say for (\ref{V1}),
is \be
\Delta s^2=\left(\Delta t-t_p\right)^2-
\Delta x^{i2}\left(1-{t_p \over \Delta t}\right)^2
.\ee
Hence Planck durations become null. This conclusion applies to other
choices of $V$ that leave the Planck scale invariant.

\section{Weighing the Options}
We presented two distinct ways to define position space in
non-linear special relativity (which are
by no means exhaustive). In the first construction 
we let the non-linear realization of the Lorentz group in momentum 
space fully fix space-time by requiring that the contraction should
remain linear, and that free field theories should have plane wave
solutions. 
The result is an energy-dependent space-time. In the second
we allow for any
non-linear representation of the Lorentz group to be introduced
{\it independently} in position and momentum space. A non-linear
contraction then carries the burden of enforcing invariance. This
precludes plane wave solutions and simple field theories and for
this reason we strongly favour the first approach.


Furthermore the first 
approach proposed in this paper simplifies enormously the
definition of a theory of gravity based on non-linear special
relativity. None of the ``doubly special'' theories have so far been
formulated to incorporate space-time curvature, and thus
gravitation. A related issue is the fate of the metric structure
of space-time in these theories, and by extension, in quantum
gravity.

It has been suggested \cite{leejoao} that non-linear realizations
of the Lorentz group (in real space) have no quadratic invariant,
and thus no metric. At high energy (or at small distances) the
concept of metric simply disintegrates. However, given that the
structure constants of the symmetry algebra are left unchanged (we
are merely changing representation), it could be that much of the
usual connection-based construction of general relativity could
still be salvaged. This program fits the second of our
constructions, but it has not, so far, been accomplished.
It may even be claimed that such an enterprise is impossible. 

Curiously, a much simpler program emerges if we adopt the
first of our constructions. In that case, the space-time transformations
remain linear, albeit energy-dependent, and so retain
a quadratic invariant. An energy-dependent  metric can still be defined,
such that in some sense, the
metric ``runs'' with the energy. This approach leads to a very
simple extension of general relativity, capable of accommodating
doubly special relativity, and is explored  in
\cite{leejoao3}. It remains to be seen how strong gravity 
phenomena -- such as black hole dynamics and Hawking's radiation -- behave
in this theory.

\section*{ACKNOWLEDGEMENTS}
We thank Chris Isham for having made us take the plunge leading to
the first approach described in this paper. JM is grateful to
Giovanni Amelino-Camelia, Jurek Kowalski-Gilkman and Lee Smolin,
for many interesting conversations and also to Max Tegmark for
a comment which inspired the field theory formulation.

\end{document}